\documentclass[prl,showpacs,aps,twocolumn]{revtex4}
\usepackage{epsfig,graphicx}

\begin{document}

\title{Evidence for In-Medium Changes of Four-Quark Condensates}

\author{
{\sc R. Thomas$^a$, S. Zschocke$^{a,b}$, B. K\"ampfer$^{a,c}$} }

\address{
$^a$ Institut f\"ur Kern- und Hadronenphysik,
Forschungszentrum Rossendorf, PF 510119, 01314 Dresden, Germany\\
$^b$ University of Bergen, Theoretical and Computational Physics Section, 
5007 Bergen, Norway\\
$^c$ Institut f\"ur Theoretische Physik, TU Dresden, 01062 Dresden, Germany
}

\begin{abstract}
Utilizing the QCD sum rule approach to the behavior of the $\omega$ meson 
in nuclear matter we derive
evidence for in-medium changes of particular four-quark condensates
from the recent CB-TAPS experiment for the reaction
$\gamma + {\rm A} \to {\rm A}' + \omega (\to \pi^0 \gamma)$ with A = Nb and
LH$_2$.
\end{abstract}

\maketitle

The chiral condensate $\langle \bar q q \rangle$ 
is an order parameter for the spontaneous breaking
of chiral symmetry in the theory of strong interaction 
(cf.\ e.g.\ \cite{Rapp_Wambach} for introducing this topic). 
The role of $\langle \bar q q \rangle$ is highlighted, e.g., 
by the Gell--Mann-Oakes-Renner relation
$m_\pi^2 f_\pi^2 \propto - \langle \bar q q \rangle$ 
(cf.\ \cite{Colangelo};
the explicit chiral symmetry breaking is essential for a finite
pion mass $m_\pi$, while the relation of the pion decay constant $f_\pi$
to $\langle \bar q q \rangle$ qualifies the latter as an order parameter)
or by Ioffe's formula $M_N \propto - \langle \bar q q \rangle$ 
for the nucleon mass (cf.\ \cite{Narison} and in particular the discussion in
\cite{Birse}).
There is growing evidence that the quark-gluon condensate
is another order parameter \cite{qg_cond}. The QCD trace anomaly related to
scale invariance breaking gives rise to the gluon condensate. 
There are many other condensates
characterizing the complicated structure of the QCD vacuum. In a medium,
described by temperature and baryon density $n$, these condensates change,
i.e., the ground state is rearranged. Since hadrons are considered as excitations
above the vacuum, a vacuum change should manifest itself as a change of the
hadronic excitation spectrum. This idea triggered widespread activities
to search for in-medium modifications of hadrons. Such in-medium modifications
of hadronic observables are found (cf.\ the lists in \cite{CB_TAPS,KEK}), and it is
timely to relate them to corresponding order parameters.

We deduce here evidence for a noticeable drop of in-medium four-quark condensates
in cold nuclear matter
from results of the recent CB-TAPS experiment \cite{CB_TAPS} for the
reaction $\gamma + A \to A' + \omega (\to \pi^0 \gamma)$. The CB-TAPS collaboration
observed the occurrence of additional low-energy $\omega$ decay strength for a Nb ($A = 93$)
target compared to a LH$_2$ ($A = 1$) target. The link of observables 
to quark and gluon condensates is established by QCD sum rules \cite{SVZ},
which are expected to be sensitive to four-quark condensates in the vector channels
\cite{Shuryak_Schafer}.
Four-quark condensate combinations which contain only left-right
helicity flipping terms (as the chiral condensate does) represent
other order parameters of chiral symmetry. 

Concentrating on the iso-scalar part of the causal current-current correlator
\cite{Narison}
\begin{equation}
\Pi^\omega(q,n) = 
\frac i3 \int d^4 x \, e^{iqx} \langle \Omega | {\cal T} j^\omega_\mu (x) {j^\omega}^\mu (0) 
| \Omega \rangle,
\end{equation}
here for the $\omega$ meson with the current 
$j^\omega_\mu = \left( \bar u  \gamma_\mu u + \bar d  \gamma_\mu d \right)/2$
and nuclear matter states $\vert \Omega \rangle$ (the symbol ${\cal T}$ means time ordering,
and $u, d$ denote quark field operators),
an operator product expansion and a Borel transformation
(cf.\ \cite{Leupold,Narison} for arguments in favor of Borel sum rules) 
of the twice-subtracted dispersion relation result in 
\begin{eqnarray}
&& \hspace*{-1cm} 
\Pi^\omega (0,n)
-\frac{1}{\pi} \int_0^\infty ds \frac{{\rm Im} \Pi^\omega(s,n)}{s} e^{-s/{\cal M}^2}
\label{sum_rule_0} \\
&& \hspace*{1cm} 
= c_0 {\cal M}^2 + \sum_{j=1}^\infty \frac{c_j}{(j-1)! {\cal M}^{2(j-1)}}, \nonumber
\end{eqnarray}
where $\Pi^\omega (0,n) = 9 n/(4 M_N)$ with the nucleon mass $M_N$
is a subtraction constant having the meaning of Landau damping
or $\omega \, N$ forward scattering amplitude, and the coefficients $c_j$ contain condensates
and Wilson coefficients; ${\cal M}$ is the Borel mass.
The first coefficients $c_j$ have been spelled out in many papers 
(cf.\ \cite{SZ_2004} for our notation, and \cite{Narison_Zakharov}
for an anomalous contribution) and are not reproduced here in full length.
$c_0 = (1 + \frac{\alpha_s}{\pi})/(8\pi^2)$
is the perturbative term. $c_1 \propto m_q^2$ is exceedingly small due to the
small current quark mass $m_q$. In $c_2$ the gluon condensate
(being less sensitive to medium effects),
some moments of the parton distribution in the nucleon 
(combined with a density dependence),
and the renormalization group invariant
combination $m_q \langle \bar q q \rangle$ (being numerically tiny) enter.
The latter fact makes the Borel sum rule insensitive to the genuine chiral condensate,
but sensitive to four-quark condensates which enter $c_3$, among other quantities
related to 
expectation values of certain traceless and symmetric twist-2 and twist-4
operators. To be specific, the flavor mixing condensates
$\frac29 \langle \bar u \gamma^\mu \lambda_A u \bar d \gamma_\mu \lambda_A d \rangle
+ \langle \bar u \gamma_5 \gamma^\mu \lambda_A u 
\bar d \gamma_5 \gamma_\mu \lambda_A d \rangle$
and the pure flavor four-quark condensates (for which we employ $u - d$ isospin symmetry;
$\gamma_\mu$ and $\lambda_A$ stand for Dirac and Gell-Mann matrices)
$\frac29 \langle \bar q \gamma^\mu \lambda_A q \bar q \gamma_\mu \lambda_A q \rangle
+ \langle \bar q \gamma_5 \gamma^\mu \lambda_A q 
\bar q \gamma_5 \gamma_\mu \lambda_A q \rangle$
enter $c_3$.
($c_4$ will be discussed below.)
Our strategy to deal with these condensates is as follows:
(i) the factorized expressions 
(which might fail badly \cite{Narison,Narison_Zakharov})
are corrected by factors $\kappa_\Omega$
(with $\Omega$ being a label of the respective four-quark condensate) 
using $\langle \bar u \gamma^\mu \lambda_A u \bar d \gamma_\mu \lambda_A d \rangle
= - \kappa_1 \frac{4}{9 \pi^2} \frac{Q_0^2}{f_\pi^2} \langle \bar q q \rangle^2$,
$\langle \bar u \gamma_5 \gamma^\mu \lambda_A u 
\bar d \gamma_5 \gamma_\mu \lambda_A d \rangle = 
\kappa_2 \frac{4}{9 \pi^2} \frac{Q_0^2}{f_\pi^2} \langle \bar q q \rangle^2$
(where $Q_0$ is a cut-off related to the $\rho - \omega$ mass splitting;
both expressions are already beyond the ground state saturation \cite{SZ_2004}),
$\langle \bar q \gamma^\mu \lambda_A q \bar q \gamma_\mu \lambda_A q \rangle = 
- \frac{16}{9} \kappa_3 \langle \bar q q  \rangle^2$,
$\langle \bar q \gamma_5 \gamma^\mu \lambda_A q 
\bar q \gamma_5 \gamma_\mu \lambda_A q \rangle = 
\frac{16}{9} \kappa_4 \langle \bar q q  \rangle^2$;
$\kappa_{1,2} = 0$ and $\kappa_{3,4} = 1$ recover the factorized terms
in the ground state saturation approximation;
(ii) expand $\kappa_\Omega$ in density \cite{Klingl_Weise}, i.e., 
$\kappa_\Omega = \kappa_\Omega^{(0)} + \kappa_\Omega^{(1)} n$,
use the known sigma term $\sigma_N$ in
$\langle \bar q q  \rangle = \langle \bar q q  \rangle_0 + \xi n$
with $\xi = \sigma_N /(2 m_q)$ \cite{Narison},
linearize the resulting expressions \cite{Oleg};
(iii) add up all contributions with their corresponding prefactors to get a common factor 
$\kappa_0 = \frac{9}{28 \pi^2} \frac{- Q_0^2}{f_\pi^2}
(\frac29 \kappa_1^{(0)} - \kappa_2^{(0)})
- \frac27 \kappa_3^{(0)}
+ \frac97 \kappa_4^{(0)}$ 
for the vacuum contribution,
$-\frac{112}{81} \pi \alpha_s \kappa_0 \langle \bar q q \rangle_0^2$,
and a common factor 
$\kappa_N = \kappa_0 + \frac{\langle \bar q q \rangle_0}{2 \xi} \left(
 \frac{9}{28 \pi^2} \frac{- Q_0^2}{f_\pi^2}
 (\frac29 \kappa_1^{(1)} - \kappa_2^{(1)})
- \frac27 \kappa_3^{(1)}
+ \frac97 \kappa_4^{(1)} \right)$ 
for the density dependent medium contribution of the mentioned four-quark
condensates (cf.\ first term in eq.~(\ref{eq.10}) below). $\kappa_0$
enters the vacuum sum rule and has to be adjusted properly with other
quantities to get the correct vacuum
$\omega$ mass, while $\kappa_N$ is subject of our further consideration.
Due to the mixing of density dependencies of $\kappa_\Omega$ and 
$\langle \bar q q \rangle$ even the accurate knowledge of $\kappa_0$
does not fix $\kappa_N$.

No density dependence of the four-quark condensates would imply $\kappa_N = 0$,
while strong density dependencies will result in a 
sizeable value of $\kappa_N$, unless the terms contributing to $\kappa_N$ cancel.
The estimate in \cite{SZ_2004} points to 
small values of $\kappa_{1,2}^{(1)}$ thus having essentially the density
dependence of the combined pure flavor scalar dimension-6 condensates
$\frac29 \langle \bar q \gamma^\mu \lambda_A q \bar q \gamma_\mu \lambda_A q \rangle
+ \langle \bar q \gamma_5 \gamma^\mu \lambda_A q 
\bar q \gamma_5 \gamma_\mu \lambda_A q \rangle$ to be constrained. 

Large-$N_c$ arguments \cite{Narison,Leupold_Nc}
favor $\kappa_N = \kappa_0$.
Previously, often the factorization 
$\langle \bar q \cdots q \bar q \cdots q \rangle \to \langle \bar q q \rangle^2$
has been used.
Here, we study explicitly, however, the role of the four-quark condensates
using the square of the genuine chiral condensate only to set the scale, as outlined above.
The integral in the l.h.s.\ of (\ref{sum_rule_0})
can be decomposed in a low-lying resonance part,
$\int_0^{s_\omega} ds {\rm Im} \Pi^\omega(s,n) s^{-1} e^{-s/{\cal M}^2}$,
and the continuum part, 
$\int_{s_\omega}^\infty ds {\rm Im} \Pi^\omega(s,n) s^{-1} e^{-s/{\cal M}^2} \equiv
- \pi {\cal M}^2 c_0 e^{-s_\omega/{\cal M}^2}$, both depending on the 
continuum threshold $s_\omega$. The quantity
\begin{equation}
m_\omega^2 (n,{\cal M}^2,s_\omega) \equiv \frac{\int_0^{s_\omega} ds \; {\rm Im} \Pi^\omega (s,n) \; 
e^{-s/{\cal M}^2}}{\int_0^{s_\omega} ds \; {\rm Im} \Pi^\omega (s,n) \; s^{-1} e^{-s/{\cal M}^2}}
\label{mass_parameter}
\end{equation}
is a normalized moment with $s$ meaning the coordinate of the center
of gravity of ${\rm Im} \Pi^\omega(s,n) e^{-s/{\cal M}^2}/s$ in the interval 
$s = 0 \cdots s_\omega$. Clearly, when
additional strength of ${\rm Im} \Pi^\omega$ at lower values of $s$
is caused by in-medium effects as observed in \cite{CB_TAPS}, 
then the center of gravity shifts to the left, i.e. $m_\omega^2$ becomes smaller.
Direct use of the count rates in \cite{CB_TAPS} (middle panel of figure 2 there)
as estimator of ${\rm Im}\Pi^\omega$ in the interval $s = 0.41 \cdots 0.77$ GeV$^2$ 
yields $m_\omega^2 (LH_2) = 0.599$ GeV$^2$ and $m_\omega^2 (Nb) = 0.568$ GeV$^2$
for ${\cal M} \sim {\cal O} (1)$ GeV.
Instead of testing the consistency of a particular model for 
${\rm Im} \Pi^\omega(s,n)$ with the sum rule, we suggest here to use the experimental
information on ${\rm Im} \Pi^\omega$ to find constraints on the QCD side 
of the sum rule. In fact, the $\omega$ decay rate 
$\omega \to \pi^0 \gamma$ is given by
$dR_{\omega \to \pi^0 \gamma} /d^4q = 
(6 d/f_\pi)^2 (\pi / [3 q^2]) (q^2 - m_\pi^2)^3\, {\rm Im}\Pi^\omega (q^2 = s)$ with
$d = 0.011$.
However, acceptance and efficiency corrections to the results of \cite{CB_TAPS}
need to be invoked and the fraction of events, where the rate
$dR_{\omega \to \pi^0 \gamma}/d M_{\pi^0 \gamma}$ is shifted to smaller
values of $M_{\pi^0 \gamma}$ (being the invariant mass of the
$\pi^0$ and $\gamma$ decay products of $\omega$)
by final state interaction of the decay $\pi^0$ in the ambient nuclear medium
\cite{Muhlich}, must be corrected for as well. 
We postpone such a quantitative and model dependent study for future work and 
consider qualitatively here the implication of the observation of \cite{CB_TAPS},
i.e., the occurrence of additional $\omega$ decay strength at  
$M_{\pi^0 \gamma} < m_\omega^{(0)}$ which translates into 
$m_\omega < m_\omega^{(0)} = 0.782$ GeV for low-momentum $\omega$
decaying in the Nb nucleus.

With (\ref{mass_parameter}) the truncated QCD sum rule (\ref{sum_rule_0}) 
for the $\omega$ meson can be arranged as \cite{SZ_2004}
\begin{eqnarray}
&& m_\omega^2 (n,{\cal M}^2,s_\omega) = \label{sum_rule}\\
&& \hspace*{-3mm} \frac{c_0 {\cal M}^2 
\left[ 1 - \left ( 1 + \frac{s_\omega}{{\cal M}^2} \right) e^{-s_\omega / {\cal M}^2} \right] - 
\frac{c_2}{{\cal M}^2} - \frac{c_3}{{\cal M}^4} - \frac{c_4}{2 {\cal M}^6}}
{c_0 \left ( 1 - e^{-s_\omega / {\cal M}^2} \right) + \frac{c_1}{{\cal M}^2} 
+ \frac{c_2}{{\cal M}^4} + 
\frac{c_3}{2 {\cal M}^6} + 
\frac{c_4}{6 {\cal M}^8} - 
\frac{\Pi^\omega (0,n)}{{\cal M}^2}}. \nonumber
\end{eqnarray}
This sum rule is to be handled as usual (cf.\ \cite{Leupold,SZ_2004}): 
determine the sliding Borel window by requiring that 
(i) the sum of the $c_{3,4}$ terms in eq.~(\ref{sum_rule_0})
does not contribute more than 10\,\% to the r.h.s., and 
(ii) the continuum part defined above does not exceed 50\,\% of the
l.h.s.\ of (\ref{sum_rule_0}) to ensure sufficient sensitivity for the resonance part; 
(iii) the continuum threshold is determined by the requirement
of maximum flatness of $m_\omega^2(n, {\cal M}^2, s_\omega)$ within the Borel window;
(iv) $m_\omega^2$ follows as average with respect to ${\cal M}^2$.

Despite the linear density expansion of the condensates entering the coefficients $c_j$, 
the sum rule (\ref{sum_rule}) is non-linear in density.
It is instructive to consider the linearized form.
Using the notation $s_\omega = s_\omega^{(0)} + s_\omega^{(1)} n$
and $c_j = c_j^{(0)} + c_j^{(1)} n$ we arrive at 
\begin{equation}
m_\omega^2 (n, {\cal M}, s_\omega^{(0)}, s_\omega^{(1)}) = R + \Delta n
\label{eq.5}
\end{equation}
with
\begin{eqnarray}
R &=& \frac{1}{N} \left\{
c_0^{(0)} {\cal M}^2 \left[1 - \left(1 + \frac{s_\omega^{(0)}}{{\cal M}^2} 
\right) E\right] \right. \\ 
&& \hspace*{9mm} \left. - \frac{c_2^{(0)}}{ {\cal M}^2} 
- \frac{c_3^{(0)}}{ {\cal M}^4}
- \frac{c_4^{(0)}}{2{\cal M}^6} \right\}, \nonumber \\
N &=& c_0^{(0)} (1 - E)
+ \frac{c_1^{(0)}}{ {\cal M}^2} 
+ \frac{c_2^{(0)}}{ {\cal M}^4}
+ \frac{c_3^{(0)}}{2{\cal M}^6} 
+ \frac{c_4^{(0)}}{6{\cal M}^8},\\
\Delta &=& \frac{1}{{N \cal M}^2} \left\{
\left[\frac{9 R}{4 M_N} + 
c_0^{(0)} E s_\omega^{(1)} ( s_\omega^{(0)} - R) \right. \right. \label{Delta}\\
&& \left. \left. - c_2^{(1)} \left(1 + \frac{R}{{\cal M}^2}\right) \right]
- \frac{c_3^{(1)}}{ {\cal M}^2} \left(1 + \frac{R}{2 {\cal M}^2}\right) \right.
\nonumber \\
&& \left.
- \frac{c_4^{(1)}}{2{\cal M}^4} \left(1 + \frac{R}{3 {\cal M}^2}\right) \right\}, 
\quad E = e^{-s_\omega^{(0)}/{\cal M}^2},  \nonumber
\end{eqnarray}
which we use for illustrative purposes. 
The quantity $R$ determines the 
vacuum properties of the $\omega$ meson; 
for the sake of estimates we can put
it equal to $m_\omega^{(0)2}$ and use ${\cal M} \sim 1$ GeV. 
$N$ contains only vacuum quantities, and $N > 0$ holds.
Therefore, the sign of the in-medium shift of $m_\omega^2$ is determined
by $\Delta$. For its estimate we note
\begin{widetext}
\begin{eqnarray}
c_2^{(1)} &=& \frac12 \left(1+ \frac13 \frac{\alpha_s}{\pi}\right) \sigma_N
- \frac{M_{N,0}}{27}
+\left(\frac14 -\frac{5}{36} \frac{\alpha_s}{\pi} \right) A_2^{u+d} M_N
-\frac{9}{16} \frac{\alpha_s}{\pi} A_2^G M_N,\\
c_3^{(1)} &=& -\frac{112}{81} \pi \alpha_s  
\frac{\sigma_N \langle \bar q q \rangle_0}{m_q} \kappa_N 
- \left(\frac{5}{12} + \frac{67}{144} \frac{\alpha_s}{\pi} \right) A_4^{u+d} M_N^3 
\label{eq.10} \\ 
&& + \frac{615}{864} \frac{\alpha_s}{\pi} A_4^G M_N^3
+ \frac14 M_N \left[\frac32 K_{u,1} + \frac38 K_{u,2} + \frac{15}{16} K_{u,g}\right] 
-\frac{7}{144} \sigma_N M_N^2,  \nonumber
\end{eqnarray}
\end{widetext}
where we include three active flavors on a 1 GeV scale;
$M_{N,0} = 0.77$ GeV is the nucleon mass in the chiral limit.
Some of the quantities in (6 - 10) are rather well known
(e.g., the twist-2 contributions), while others are individually
less accurately fixed. 
We use for our evaluations
$\langle \bar q q \rangle_0 = (-0.245 \,{\rm GeV})^3$,
$\sigma_N = 0.045$ GeV,
$\alpha_s = 0.38$,
$Q_0 = 0.15$ GeV,
$f_\pi = 0.093$ GeV,
$A_2^{u+d} = 1.02$,
$A_2^G = 0.83$,
$A_4^{u+d} = 0.12$,
$A_4^G = 0.04$,
$K_{u,1} = -0.112$ GeV$^2$,
$K_{u,2} = 0.11$ GeV$^2$,
$K_{u,g} = -0.3$ GeV$^2$
\cite{SZ_2004} to get 
$c_2^{(1)} \approx 0.17$ GeV and 
$c_3^{(1)} \approx 0.2 (\kappa_N - 0.7) \,{\rm GeV}^3$
to obtain finally
\begin{equation}
\Delta \approx (4 -\kappa_N) \frac{0.03}{n_0} \,{\rm GeV}^2,
\label{eq.11}
\end{equation}
where we employ 
$s_\omega^{(0)} = 1.4$ GeV$^2$,
$s_\omega^{(1)} = - 0.15 n_0^{-1}$ GeV$^2$ 
(with $n_0 = 0.15$ fm$^{-3}$ as nuclear saturation density)
from an evaluation basing on (\ref{sum_rule})
and neglect the $c_4$ term for the moment being. 
To probe the uncertainty caused by less constrained quantities
in (6 - 10) we assign 
$N$, $R$, $c_0^{(0)}$, $c_2^{(1)}$, and $c_3^{(1)}$
(the term in front of $\kappa_N$ and the remainder separately) 
the large uncorrelated variations of $\pm 10$\,\% and arrive at 
$\Delta = (2.8 \cdots 5.3 - \kappa_N)(0.023 \cdots 0.035) n_0^{-1}$ GeV$^2$.
In essence
to achieve a negative value of $\Delta$ and thus the experimentally observed 
\cite{CB_TAPS} dropping
of $m_\omega$ in medium, a sufficiently large value of $\kappa_N$ is required,
as evidenced by eqs.~(\ref{eq.5}) and (\ref{eq.11}).
Thereby, the term $\propto c_3^{(1)}$ provides a counterbalance to the
large Landau damping \cite{Hoffmann} (first term in (\ref{Delta})).
(For the $\rho$ meson the Landau damping term is nine times smaller \cite{Hoffmann}, 
resulting in an always negative shift parameter conform with the
dropping $\rho$ mass scenario in \cite{Hatsuda_Lee} and 
in qualitative agreement with the Brown-Rho scaling \cite{BR}.)

Indeed, the evaluation of the complete
sum rule (\ref{sum_rule}) requires for the described parameter set $\kappa_N \ge 4$
to have $m_\omega^2 (n > 0) < m_\omega^{(0) 2}$,
see Fig.~1. In other words, the above mentioned four-quark condensates
must change drastically in the nuclear medium.
With the above quoted parameters this translates into
the huge amount of more than a 50\,\% drop of the combined four-quark condensates
at nuclear matter saturation density when
relying on the linear density expansion up to such density.
(The experiment \cite{CB_TAPS} probes actually densities $\sim 0.6 n_0$.)
Phrased differently, the density dependence of the $c_3$ term must be stronger
than the simple factorization allows.

To have some confidence in our estimate, the influence of $c_4$ must be evaluated.
An order-of-magnitude estimate utilizing ground state saturation would yield
$c_4 = (5.48 + 0.122 n/n_0) \times 10^{-5}$ GeV$^8$ when considering only the first seven
mass dimension-8 scalar condensates \cite{first_seven} 
and the two twist-2 condensates \cite{twist_2}. 
The corresponding value of $c_4^{(0)}$ is substantially smaller than
the standard estimates related to the $\tau$ decay:  
\cite{Dominguez} quotes
$(-7 \cdots + 4) \times 10^{-3}$ GeV$^8$ pointing to some uncertainty
also of $c_4^{(1)}$. We consider, therefore, $c_4$ as parameter and study its
impact on the sum rule, as illustrated in Fig.~1.

Experimentally, a much stronger drop of the $\omega$ meson mass squared
is found \cite{CB_TAPS} than the in-medium change of $m_\omega^2$ exhibited
in Fig.~1. The conservative estimate of 
$\kappa_N > 4^{+ 0.7}_{-0.7} - 2.8^{+0.2}_{-0.4} \times 10^3 {\rm GeV}^{-8} n_0 c_4^{(1)}$
obtained from an evaluation of the sum rule (\ref{sum_rule}) expresses a condition
for a decreasing value of $m_\omega^2$ in the nuclear medium
(the indicated variation of the numbers arise from an assumed uncertainty of
$m_\omega^{(0)\,2}$ by $\pm 10$\,\% which should reflect the approximate 
character of the vacuum sum rule).   
In other words, as long as $c_4^{(1)} < 1.7^{+0.3}_{-0.6} \times 10^{-3} n_0^{-1}$ GeV$^8$
a finite positive value of $\kappa_N$ is required, i.e.,
a noticeable density dependence of the combined four-quark condensates.
Note that this statement is independent of a model for ${\rm Im}\Pi^\omega$;
it bases only on the observation that $m_\omega^2$ must become smaller
if additional strength of ${\rm Im}\Pi^\omega$ below $m_\omega^{(0) 2}$
occurs in the medium, as observed in \cite{CB_TAPS}.

Finally, we mention that many more four-quark condensates 
enter other sum rules in different combinations. For instance, the investigation of the
three coupled sum rule equations for the nucleon \cite{Furnstahl}
points to some cancellations among the four-quark condensates
when using the estimates from \cite{Tuebingen}. 
(The results of \cite{Tuebingen} can not be employed directly for our
$\omega$ sum rule since the flavor-mixing four-quark condensates are delivered
in color combinations suitable only for the nucleon sum rule.)
Furthermore, a crucial point is that the genuine chiral condensate is not suppressed in the
nucleon sum rule, while in the $\omega$ sum rule it is. 

In summary we argue that the recent CB-TAPS experiment \cite{CB_TAPS} 
implies a noticeable drop (more than 50\% when extrapolating to nuclear
saturation density and truncating the sum rule beyond mass dimension-6) 
of a certain combination of four-quark condensates.
Four-quark condensates are fundamental quantities, among others, characterizing
the non-perturbative QCD vacuum. Specific four-quark condensates,
changing under chiral transformation, represent further important order
parameters for chiral symmetry restoration.
Clearly, also other channels, besides the omega meson considered here,
must be studied to gain more information on these particular four-quark condensates.

\begin{acknowledgments}
{\bf Acknowledgments:} 
The authors thank R. Hofmann, S. Leupold, W. Weise
for stimulating discussions and V. Metag and D. Trnka for informing us 
on the status of the CB-TAPS experiment. The work is supported by BMBF 06DR121
and GSI-FE.
\end{acknowledgments}

\begin{figure}[h]
\includegraphics[width=6cm,angle=-90]{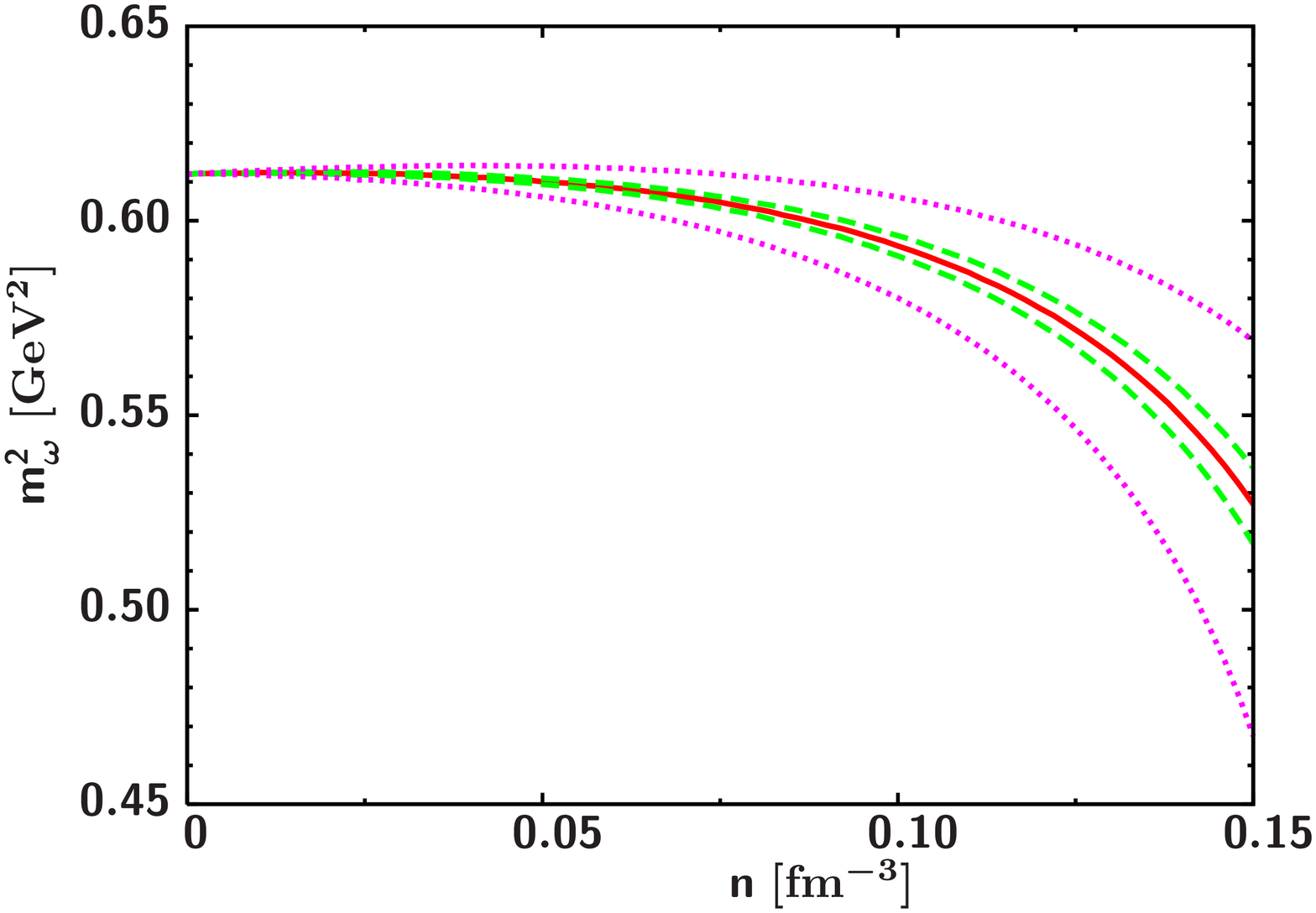}
\caption{\label{fig1}
The mass parameter $m_\omega^2$ defined in eq.~(\ref{mass_parameter})
and averaged within the Borel window as a function of the baryon density
for $\kappa_N = 4$ and $c_4 = 0$ (solid curve). 
Note that the parameter $m_\omega^2$ coincides only in zero-width approximation 
with the $\omega$ pole mass squared; in general it is a normalized
moment of ${\rm Im}\Pi^\omega$ to be calculated from data or models.
The sum rule eq.~(\ref{sum_rule}) is evaluated
as described in the text with appropriately adjusted $\kappa_0$.
Inclusion of $c_4^{(0)} = {\cal O} (\pm 10^{-3})$ GeV$^8$
requires a readjustment of $\kappa_0$ in the range $1 \cdots 5$
to $m_\omega^{(0) 2}$. A simultaneous change of $\kappa_N$
in the order of 20 \% is needed to recover the same density dependence as given by
the solid curve at small values of $n$.
The effect of a $c_4^{(1)}$ term is exhibited, too
($c_4^{(1)} = \pm 10^{-5} n_0^{-1}$ GeV$^8$: dashed curves,
$c_4^{(1)} = \pm 5 \times 10^{-5} n_0^{-1}$ GeV$^8$: dotted curves;
the upper (lower) curves are for negative (positive) signs).}
\end{figure}

\end{document}